\documentclass[11pt,oneside]{article}
\usepackage{graphicx}
\begin{document}
 
\title{{\bf The Age--Metallicity Relation in the Thin Disk of the Galaxy}}
\author{{\bf  V.~A.~Marsakov, V.~V.~Koval', T.~V.~Borkova and M.~V.Shapovalov}\\
Institute of Physics, Southern Federal University,\\
Rostov-on-Don, Russia\\
e-mail:  marsakov@ip.rsu.ru, koval@ip.rsu.ru, borkova@ip.rsu.ru}
\date{accepted \ 2011, Astronomy Reports, Vol. 55 No. 8, P.667-682}
\maketitle

\begin {abstract}
HST trigonometric distances, photometric metallicities, isochronic ages 
from the second revised version of the Geneva--Copenhagen survey, and 
uniform spectroscopic $Fe$ and $Mg$ abundances from our master catalog are
used to construct and analyze the age--metallicity and age--relative $Mg$
abundance relations for stars of the thin disk. The influences of 
selection effects are discussed in detail. It is demonstrated that the 
radial migration of stars does not lead to appreciable distortions in the
age dependence of the metallicity. During the first several billion 
years of the formation of the thin disk, the interstellar material in 
this disk was, on average, fairly rich in heavy elements 
($\langle [Fe/H]\rangle \approx -0.2$) and poorly mixed. However, the 
metallicity dispersion continuously decreased with age, from 
$\sigma_{[Fe/H]}\approx 0.22$  to $\approx 0.13$. All this time, the 
mean relative abundance of $Mg$ was somewhat higher than the solar 
value ($\langle [Mg/Fe]\rangle \approx 0.1$). Roughly four to five 
billion years ago, the mean metallicity began to systematically 
increase, while retaining the same dispersion; the mean relative $Mg$ 
abundance began to decrease immediately following this. The number of 
stars in this subsystem increased sharply at the same time. 
These properties suggest that the star-formation rate was low in 
the initial stage of formation of the thin disk, but abruptly 
increased about four to five billion years ago.

\end {abstract}

\section*{Introduction}

The time for the existence of the thin-disk subsystem
is comparable to the age of the Galaxy itself.
Therefore, ongoing elemental-synthesis processes
during this period should lead to an appreciable
enhancement of the general heavy-element abundance
in young stars of this subsystem. As a result,
we would expect the thin disk to display a well defined
age--metallicity dependence, with the highest
metallicities being observed in the youngest stars.
However, numerous investigations into this question
have left the existence of such a dependence open.

The first systematic studies of the relationship between
age and metallicity in the Galactic disk were
those carried out by Twarog~[1] for $\sim1000$ field stars,
based on data obtained in the mean-filter Str\"omgren
photometric system, which made it possible to carry
out a three-dimensional classification of these stars.
The results were consistent with expectations, since
they demonstrated that, beginning with the oldest
stars of the subsystem, the metallicity increased uniformly
with decreasing age. Str\"omgren~[2] underscored
another aspect of the age--metallicity relation
found by Twarog: the very small dispersion of the
metallicity $\sigma_{[Fe/H]}$ for stars of a given age, which
only slightly exceeded the scatter due to observational
uncertainties.

Subsequently, both of these conclusions evaporated.
Based on $uvby$ photometry for  $\sim 5500$ F2--\,G2
dwarfs of the disk, Marsakov~et\,al.~[3] showed, using
kinematics as a statistical indicator of the age of a
stellar group, that the age--metallicity dependence is
two-dimensional, with the following main properties:
a large scatter in the metal contents of old stars, and
a lowering of the metallicity scatter together with an
increase in the mean metallicity with decreasing age.

Based on more reliable spectroscopic measurements
of the $Fe$ abundances and isochronic ages for
roughly 200 ~stars of the disk, Edvardsson~et\,al.~[4]
established that the real scatter in the metallicities
for stars of a single age was appreciably higher than
the observational uncertainties. Simultaneously, attention
was turned to the existence of very old stars
with high metallicities in the subsystem. Edvardsson~et\,al.~[4] 
also asserted that, if an age dependence
of the mean metallicity in the disk exists, it is very
weak. However, all these early studies were subject
to $a$ $priori$ selection effects, due to the limited nature
of samples of stars with temperatures below spectral
type G2, leading to an obvious exclusion of possible
old, metal-rich stars, whose turn-off points lie in the
region of lower temperatures, according to theoretical
calculations (see Fig.~2a and the sloped dashed line,
corresponding to $T_{eff} = 5800$ K).

In recent studies based on new $uvby$ photometry
data and satellite astrometric data for several
thousand disk stars ($5-7$), it has become possible
to include stars to the end of spectral type G, and
even later-type stars. The ages in [5] are based
on theoretical evolutionary tracks constructed using
an algorithm that minimized the difference between
the observed and theoretical magnitudes and effective
temperatures for each star. This algorithm excludes
the main-sequence turn-off, where the isochrones
admit the existence of stars with the same atmospheric
parameters but different ages. The studies
[6,\,7] represent different versions of the Geneva--Copenhagen 
survey (the main version, and a later
version corrected for systematic errors in the stellar
parameters). These ages were estimated using
theoretical isochrones based on Bayesian theory~[8].
This method takes into account systematic effects,
and enables more reliable estimation of stellar ages
based on probability distribution functions. All the
scales are consistent with ages obtained based on the
chromospheric activity of the stars.

Numerous studies have shown the existence of
very old super-metal-rich stars, and simultaneously
asserted that all features on age--metallicity plots giving
rise to the observed correlation are due to various
selection effects. In particular, it is suggested in~[6]
that the observed higher mean metallicities for stars
younger than three billion years have two artificial origins:
first, high-temperature limits imposed on samples
and, second, the presence of distant, hot, highluminosity
stars in samples selected based purely on
apparent magnitude, whose ''supermetallicities'' are
due exclusively to ''overcorrection'' of the photometry
data for interstellar reddening.

However, following the traditional method for determining
isochronic ages, Pont and Eyer~[9] showed
that the non-linear division of the isochrones in the
Hertzsprung--Russel diagram leads to statistical selection
effects when determining isochronic stellar
ages, which artificially ascribe them large ages. Analyzing
various errors in atmospheric parameters and
isochronic ages of stars in~[4], Pont and Eyer~[9]
concluded that the identification of old, metal-rich
stars ($t > 5$ billion years, $[Fe/H] \sim 0.0$) and young,
metal-poor stars ($t < 5$ billion years, $[Fe/H] < -0.5$)
is erroneous. They explain that there are systematic
errors in the ages of these stars as a consequence
of the very simple method used to derive ages from
theoretical isochrones, and assert that application of a
correct statistical method corrected for this selection
effect would indicate that the Geneva--Copenhagen
catalog data actually indicate a monotonic growth
in the metallicity in the Galactic disk with approach
toward the current epoch, with dispersions 
$\sigma_{[Fe/H]} < 0.15$ for stars of the same age

These conclusions are in agreement with the results
of Rocha-Pinto~et\,al.~[10,\,11], who used data
on the chromospheric activity of several hundred
nearby main-sequence dwarfs to estimate their ages,
and found a significant trend for the mean metallicity
with a fairly small dispersion for stars of a single age
($\sigma_{[Fe/H]} \sim 0.12$). (Note, however, that the photometric
and spectroscopic metallicities in that study were
very poorly correlated.) Subsequently, Rocha-Pinto
et\,al.~[12] suggested that evidence for the existence
of an age--metallicity relation in the thin disk was
also provided by the observed decrease in the mean
metallicity with both increasing and decreasing mean
orbital radius of stars that are currently located near
the Sun. If the mean radius of a star's orbit is an indicator
of the Galactocentric distance of its birth~[4,\,13],
the increase in the difference between the mean orbital
radii of stars and the radius of the solar orbit with
increasing stellar age observed for nearby field stars
indicates that the mean metallicity falls off with age in
the disk. (Note that~[14] earlier explained the observed
break in the dependence of the mean metallicities
of stars of this subsystem on their mean radii as a
consequence of a decrease in metallicity with age in
the Galactic disk.)

Reid~et\,al.~[15] composed a representative sample
of nearby stars based on Hipparcos data for studies of
the age--metallicity relation. One characteristic feature
of the sample is that it includes only stars whose
main-sequence lifetimes are comparable to or exceed
the age of the thin disk (i.\,e., stars with $M_{V} > +4^{m}$).
The sample is limited from below by the absolute
magnitude, $M_{V} < +6^{m}$, which makes it complete
for F5--K0 dwarfs located within $\approx 30$~pc. They
determined the metallicities of the stars using the
calibration of~[16], based on Str\"omgren photometric
data; as the authors themselves note, the resulting
metallicities were systematically higher than those
in~[4] by $ \Delta [Fe/H]\approx 0.1$. It is striking that 
the ages they found based on the Yale isochrones~[17] are fairly
well correlated with the ages of~[4], but appear completely
uncorrelated with the ages of~[6]. As a result,
Reid~et\,al.~[15] concluded that the mean metallicity in
the solar neighborhood increases from $[Fe/H]\approx -0.3$
for an age of ten billion years to $\approx +0.15$ at the current
epoch, although the dispersion of the metallicity at
any given epoch is fairly high.

Haywood's~[18,\,19] analysis of the kinematics of
nearby stars showed that the orbital elements of the
most metal-rich and metal-poor stars currently located
in the solar neighborhood suggest that most
of them were born in inner and outer regions of the
Galactic disk, respectively. As a result, as Haywood
notes, we actually only see an absence of metal-poor
stars younger than two billion years and an increase
in the metallicity dispersion with age in the observed
age--metallicity diagram. He concludes that thin disk
stars at the solar Galactocentric distance were
born within a narrow range of metallicities ($-0.2 <
[Fe/H] < 0.2$), supporting the hypothesis that the
chemical evolution in the Galaxy primarily preceded
the formation of the disk subsystem. Simultaneously,
Roskar~et\,al.~[20] numerically modeled hydrodynamical
processes in the formation of the Galactic disk,
showing that the virtual absence of a slope and the
large scatter in observed age–-metallicity diagrams
and the comparatively small number of young, metal-poor
stars in the solar neighborhood could be explained
by a radial migration of stars.

However, Karatas et\,al.~[21] suggest that it is not
possible to draw firm conclusions about the existence
of an age–-metallicity dependence in the thin disk for
$t > 3$ billion years, due to the scatter in the metallicities
for stars of the same age and the existence of old,
metal-rich stars, supporting the conclusions of [5,\,6].
They reached this conclusion based on their analysis
of a sample containing more than four hundred stars,
which they identified with the thin-disk subsystem
using the criterion of~[22], applying information about
both the kinematics and metallicities of the stars.
They used only dwarfs near the turn-off point and
subgiants, since their isochronic ages can be determined
with high accuracy.

As the existence of monotonic metallicity variations
with age in the thin disk remains unproven, our
goal here is to carry out a detailed analysis of all possible
selection effects capable of distorting the form of
the observed age--metallicity diagram, and to investigate
variations in the character of this diagram based
on stars born at different Galactocentric distances.
To improve the reliability of the results, we consider
spectroscopic $Fe$ and $Mg$ abundances together with
photometric metallicities, as well as independent age
estimates.

\section*{OBSERVATIONAL DATA}

Holmberg et\,al.~[7] recently published a revised
version of the Geneva--Copenhagen survey, which
includes appreciably different stellar metallicities and
ages than in the previous edition~[6,\,23]. Therefore, we
used this revised version, which contains the atmospheric
parameters, ages, metallicities, and kinematics
of 14 000 F--K dwarfs, as the main source of data
for our study, supplementing this with our master
catalog of spectroscopic $Fe$ and $Mg$ abundances in
867~stars with accurately known parallaxes~[24]. The
atmospheric parameters and metallicities in the former
catalog were based on $uvby\beta$ photometric indices
corrected for interstellar reddening. Figures~1a,\,b depict
the stars present in both catalogs used. In each
panel, we present lines corresponding to full agreement
between the quantities plotted, together with
linear fits to the data. For the vast majority of these
stars, the photometric metallicities and effective temperatures
coincide with the spectroscopic values from
our master catalog within the errors. Neither of the
photometric parameters dislay any overall systematic
deviations, only small inclinations of the linear fits relative
to the median lines. The metallicities of the stars
that are richest in heavy elements are slightly underestimated,
by $\Delta[Fe/H] \leq 0.1$, while the effective
temperatures of the hottest stars are underestimated
by $\Delta T_{eff} \leq 50$~K. Such small discrepances will not
lead to any appreciable increase in the uncertainties in
the derived ages for the majority of the catalog stars.

The distances for approximately $75\,\%$ of the stars
in~[7] are based on Hipparcos trigonometric parallaxes;
only parallaxes with uncertainties below $13\,\%$
were included. We used photometric distances for
stars without parallaxes or stars whose parallaxes
have larger errors. The cited accuracy for these distances
is also about $13\,\%$. Analysis of the uncertainties
in the trigonometric parallaxes showed that
photometric distances in the catalog were adopted
primarily for distant stars, whereas $98\,\%$ of stars within
70\,pc of the Sun have trigonometric distances.

Holmberg~et~al.~[7] calculated the most probable
ages for the stars based on the theoretical isochrones
of~[25,~26], taking into account the uncertainties in
the effective temperature, absolute magnitudes, and
metallicities. Unfortunately, the variability of the
speeds with which the stars moved along the evolutionary
tracks was automatically also taken into
account, leading to some distortion of the ages of
stars located near the turn-off point, since stars in
these sections evolve comparatively rapidly, and the
probability of finding them in this stage is very low.
As a result, stars located near the evolutionary stage
corresponding to the depletion of the last percent of
hydrogen in their cores just prior to their collapse were
ascribed ages that were too young.

This age distortion affected only stars with metallicities
such that neighboring isochrones in the region
of the turn-off point have similar shapes and intersect.
This effect can be seen in the age.absolute magnitude
diagram constructed for all stars in the catalog
in Fig.~1 of~[27] (although this plot was constructed
for data from the previous version of the catalog~[23],
its character remains the same). Two separate sequences
can be clearly distinguished in this diagram,
which close up near two billion years due to the
appreciable motion away from the main-sequence of
the turn-off points for younger isochrones with higher
heavy-element contents, while the gap between the
sequences is filled near ages of $t > 5.5$ billion years
by older stars that have evolved beyond their turn-off
points. We nevertheless considered it to be possible
to use these data, since the uncertainty in assigning
ages for stars near the turn-off point is less than a
billion years, as can be seen in the figure cited above.
Koval'~et\,al.~[27] emphasize that their procedure for
calculating the ages does not distort the general trend
on the age.metallicity diagram. The analysis of the
age uncertainties of [7] showed that mean uncertainties
of $t < \pm 2$ billion years are achieved for $\approx 85\,\%$ 
of the stars, while uncertainties of $t < \pm3$ billion yeras
are achieved for $\approx 90\,\%$ of the catalog stars.

Radial velocities at different epochs were obtained
in~[7] as part of the CORAVEL project, while the
proper motions were taken primarily from the Tycho-2 
catalog. The resulting uncertainty in the stellar velocity
components is $\pm1.5$~km\,s$^{-1}$. The orbital parameters
in~[7] were calculated based on the Galactic gravitational
potential of [28], and assumed a solar Galactocentric
distance of 8\,kpc and a rotational velocity of
the thin disk at the solar distance of 220~km\,s$^{-1}$.

We used the method proposed in~[27] to select
stars for our analysis for which the probability of
membership in the thin disk exceeds the membership
probability for the thick disk. This method uses
the dispersions of each of the three spatial velocity
components obtained in~[27] and the mean rotational
velocity of the two subsystems at the solar Galactocentric
distance. Verification based on our master
catalog of spectroscopic $Fe$ and $Mg$ abundances for
nearby stars~[24] shows that this kinematic criterion
is in very good agreement with chemical criteria that
indicate that the majority of thin-disk stars selected
in this way have low ratios $[Mg/Fe] < 0.25$ and high
ratios $[Fe/H] > -0.5$ (see, in particular, ~[29,\,30]).

After removing binary stars, very evolved stars
($\delta M_{V} > 3^{m})$, and stars with uncertain ages 
($\epsilon(t) >\pm 3$ billion years), the remaining sample 
had 5805~presumed single thin-disk stars. (The resulting mean
age uncertainty for stars in this sample was $\epsilon(t) =
1.0$ billion years). Figure~1c plots the effective temperature
versus absolute magnitude for this sample
(for comparison, the small symbols show all thin disk
stars from the input catalog). It is possible to
obtain trustworthy age estimates only for stars with
absolute magnitudes $M_{V}\leq 4.6^{m}$: stars with lower
luminosities fall in the region where the isochrones
are densely packed and the age uncertainties become
very large. The stars in the resulting thin disk
sample cover a fairly wide range of effective temperature,
$\approx(5400-7000)$\,K, which obviously includes
both some of the oldest and some very young stars of
this subsystem.

The analysis of~[27] showed that the left wing of
the distribution of the distance from the Sun ($d$) for
the sample stars can be fit well by a power law, with
the index equal to two within the errors for $d < 60$\,pc,
as is expected if the distribution of stars is uniform
in the studied volume. To eliminate selection effects
associated with the different depths of the survey for
stars of different metallicities and temperatures, it is
desirable to restrict the sample to a distance from the
Sun of 60\,pc, within which the sample can be considered
to be complete. However, taking into account
the fact that the maximum of the distance distribution
is observed at a distance of $\approx 70$\,pc and the number of
nearby F--G stars suitable for our statistical analyses
is limited, we decided to restrict our sample to this
slightly greater distance (leading to an increase in the
sample volume by about $20\,\%$). As a result of this
restriction, we simultaneously minimize the errors
associated with correcting the photometric indices for
interstellar reddening, since reddening is negligible
out to this distance. Moreover, the trigonometric
parallaxes of nearly all stars within this distance are
independent of photometric measurements. The final
sample contains 2255~nearby thin-disk stars.

We used data from our master catalog of reference
spectroscopic $Fe$ and $Mg$ abundance measurements~[24] 
to identify possible selection effects due
to deriving the metallicities from photometric data,
and also to study the age dependence of the relative
abundances of $\alpha$ elements. This catalog is a
collection of virtually all $Mg$ abundances for dwarfs
and subgiants in the solar neighborhood derived via
synthetic modeling of high-dispersion spectra published
up to January 2004. The internal accuracy of
the relative $Mg$ abundances for metal-rich ($[Fe/H] > -1.0$) 
stars was $\epsilon([Mg/Fe]) = \pm 0.05$, while the
corresponding internal accuracy for the $Fe$ abundances
was $\epsilon([Fe/H]) = \pm 0.07$. The distances and
spatial velocities of the stars were calculated using
data from large, modern, high-accuracy catalogs. We
used trigonometric parallaxes with uncertainties less
than $25\,\%$, or, if these weren't available, photometric
distances based on $uvby\beta$ photometry. The catalog
initially contains some selection effects distorting the
relative numbers of stars with different metallicities
and temperatures~[24]; the subsample that is within
40\,pc of the Sun is fairly representative of F--G dwarfs
of the thin disk. Nevertheless, to retain a sufficient
number of stars with spectroscopic chemical compositions
four our analysis, we decided to restrict
the sample to a somewhat larger distance (70\,pc),
since it was only used as a supplementary sample
in our study. After applying the probability criterion
of~[27] and excluding stars with large age uncertainties
($\epsilon(t) > \pm 3$ billion years), the sample included
only 220 thin-disk stars.

\section*{ANALYSIS OF SELECTION EFFECTS 
IN THE AGE--ME-TALLICITY DIAGRAM}

Figure~2a presents the age–-metallicity diagram
for the stars in our thin-disk sample from~[7]. The
small number of stars with ages exceeding 12 billion
years are not shown, since such great ages are
likely to be overestimated, as was noted by Holmberg~et\,al~[7]. 
(Stars with spectroscopic metallicities are
shown by big, hollow circles.) The large, hollow
circles show the mean metallicities for stars in ten
narrow age ranges, each containing 225~stars. The
uncertainties in the mean values are comparable to
the sizes of the big, hollow circles. The smooth
curve is a third-order polynomial fit to all the stars,
which traces the behavior of the mean points along
the entire horizontal axis well and effectively smooths
the desired age dependence of the metallicity. Upper
and lower envelopes at the $5\,\%$ level are also shown.

For comparison with independent age estimates,
Fig.~2b presents analogous diagrams for stars with
ages from~[5,\,11]. Both of these studies used $uvby\beta$
photometry to determine $[Fe/H]$, but our analysis
showed that their values differ appreciably from those
derived from later spectroscopic and photometric data,
and we have accordingly used the metallicities
from~[7]. The ages in the former study were obtained
from theoretical isochrones, and in the latter study
from the chromospheric activity of the stars. Since
there is no kinematic data for the reliable identification
of thin-disk stars within 70\,pc of the Sun in these
studies, we identified stars coinciding with those in
our sample based on their HD numbers. This left
1078 stars from~[5], and only 206 from~[11]. Comparisons
show that the distribution of the stars and the
general run of the age dependences of the metallicity
in the lower diagram are overall consistent with the
data in the upper diagram; therefore, we will further
focus on the latter data.

Thus, Fig.~2a shows that the largest scatter in
$[Fe/H]$ is observed for old stars of the Galactic disk,
while this scatter is appreciably smaller for young
stars due to the obvious deficit of metal-poor stars
among them (the lower-left corner of the diagram is
nearly empty). This is explained in~[6] as an effect
of the limitation of the input Geneva.Copenhagen
survey at high temperatures due to the restriction on
the temperature index $b-y$. In the opinion of~[6],
the difference in the isochronic ages of metal-rich
and metal-poor stars with the same temperatures can
explain the observed slope of the lower envelope in
the diagram. The effect of limiting the sample at
high temperatures somewhat distorts the real age--
metallicity diagram.

The dashed line in Fig.~2a is based on the theoretical
isochrones used in~[7] for $T_{eff} = 7000$~K, which
corresponds to the boundary to the right of which
stars hotter than this temperature are not included
in the sample. This temperature corresponds to the
upper temperature limit for our sample. The region
to the left of this line contains mainly cooler stars
that have not moved far from their zero age main
sequences, as well as small numbers of massive stars
in the subgiant stage. As can be seen in the diagram,
to some extent, the high-temperature limit of the
sample indeed leads to a small inclination in the age
dependence of the mean metallicities for ages less
than $\approx 1.5$ billion years. (We already noted above
that the lower temperature limit of the sample of
roughly the solar temperature [see the right dashed
line] led to the exclusion of the oldest metal-rich stars
from the sample~[1], bringing about the conclusion of
a monotonic increase in metallicity with age in the
disk subsystem of the Galaxy.) However, as we can
see in the diagram, the age dependence of the mean
metallicity is determined by the relative numbers of
stars with different metallicities but the same age,
not by this nearly vertical line, to the right of which,
as before, we observe a deficit of young, metal-poor
stars. In particular, we can see that, with increasing
temperature, the densities of stars in the diagram
with $[Fe/H] < -0.1$ begin to grow sharply at greater
distances from the line $T_{eff} = 7000$~K as the metallicity
decreases. This could come about due to the
existence in the thin disk of main-sequence turn-off
points for stars with different metallicities, assuming
that metal-poor stars have been born in relatively
small numbers in recent times. We will now verify this
hypothesis.

\section*{Main-Sequence Turn-off Points for Stars
with Different Metallicities}

The left panels in Fig.~3 present $T_{eff}-M_{V}$ plots for
stars within 70\,pc of the Sun in four fairly narrow
metallicity ranges; the right panels show histograms
of the temperature distributions for these same stars.
The line segments in the histograms schematically
denote the behavior of the left envelopes of these distributions.
The progressive decrease in the numbers
of stars in the right-hand sections of the distributions
is due to the limited depth of the survey of the input
catalog (magnitudes $< 8.5^{m}$), as well as selection
effects associated with uncertainties in the age determinations
[Fig.~1c].) All the metal-poor groups
($[Fe/H] < 0.0$) display breaks in their envelopes --
inflections. While the increase in the number of stars
with decreasing temperature to the left of the ''inflection
point'' essentially corresponds to a Salpeter mass
distribution, the number of stars grows with decreasing
temperature in a jump-like fashion to the right of
this point. Note that the temperatures of these points
are far to the left of the edge of the diagram for all the
metal-poor groups. This means that the sharp deficit
of hotter (and therefore younger) metal-poor stars is
not associated with the high-temperature limit for the
sample, but instead with the existence of a minimum
age for the majority of stars of a given metallicity in
the thin disk; i.\,e., with the existence in the thin disk
of main-sequence turn-off points for metal-poor field
stars.

Theoretical isochrones from~[17] passing through
the ''inflection points'' in the right-hand panels are
shown in the three left-hand panels of Fig.~3 (for
comparison, isochrones for a higher age are also
shown to the right of these curves). Isochrones for
the most metal-rich group are not presented, since
the ''inflection point'' is not clearly distinguished in
this case. With growth in the metallicity, the position
of the turn-off point shifts toward higher temperatures
(younger ages), remaining within the studied
temperature range for stars with $[Fe/H] < 0.0$. The
small number of stars that are hotter than the turnoff
point are probably blue stragglers or stars born in
small numbers after the formation of the main mass of
stars of the given metallicity in the Galactic disk.

\begin{table}
\centering
\caption{%
   Kinematic parameters of four groups of nearby thin-disk stars}
\begin{tabular}{|l|c|c|c|c|}
\hline
\multicolumn{1}{|c|}{\bf Parameter} & 
\multicolumn{2}{|c|}{\bf $t < 2$ billion years} &
\multicolumn{2}{|c|}{\bf $t > 5$ billion years}\\
\cline{2-5}
&{\bf $[Fe/H] > 0.0$}&{\bf $[Fe/H] < -0.3$}&{\bf $[Fe/H] > 0.0$}
&{\bf $[Fe/H]< -0.3$}\\
\hline

  ($V_{LSR}$),\,km~$s^{-1}$ &$26\pm1$&$32\pm2$&$44\pm2$&$48\pm2$ \\
  $\sigma_{V}$,\,km~$s^{-1}$ &$11\pm1$&$10\pm2$&$19\pm2$&$21\pm2$ \\    
                        $N$ &   197  &   33   &   133  &   159   \\   

\hline
\end{tabular}
\end{table}

\section*{Do Old Metal-Rich Stars Exist?}

In spite of the fact that, according to theoretical
isochrones, the positions of a fairly large number of
stars with the solar metallicity on the Hertzsprung–
Russell diagram indicate old ages, the existence of
old, metal-rich stars was subject to doubt in~[9].
Pont and Eyer~[9] arrived at this conclusion after
discovering that the spectroscopic data in the sample~[4] 
for most metal-rich stars older than five billion
years display substantial discrepancies in one of the
following parameters: their trigonometric distance,
photometric temperature, photometric metallicity, or
isochronic ages from~[6]. They suggest that these discrepancies
led to an artificial enhancement in the ages
of stars located in the region of enhanced isochrone
density, so that the apparent existence of old, metal-rich
stars is an artefact.

However, Fig.~2a shows that our sample of thin disk
stars with spectroscopic Fe abundances and
stars of the Geneva--Copenhagen survey from~[7]
with revised ages likewise demonstrate a large number
of such stars. Of course, a reduction in the
metallicities in~[7] for the richest stars observed in
Fig.~1a could testify to a selective overestimation of
the ages of the oldest metal-rich stars. To verify this
hypothesis, the hollow circles in Fig.~1 show stars
with ages exceeding five billion years. Most old stars
lie below the diagonal line; i.\,e., their $[Fe/H]$ values are
all reduced, independent of their metallicity (within
the uncertainties).

This distortion leads only to a small statistical
overestimation of the ages of old stars of any metallicity.
Our analysis shows that taking into account
the deviations between the spectrosopic and photometric
metallicities and effective temperatures in
Figs.~1a and~1b does not appreciably change the relative
number of stars in the upper-right corner of the
age--metallicity diagram. The chromospheric ages
from~[11] in Fig.~2b also display some number of such
stars. In other words, independent spectroscopic
atmospheric parameters, $Fe$ abundances, and ages
confirm the existence of an appreciable number of old,
metal-rich stars in the thin disk, near the Sun.

The effect of unresolved binarity of some stars
could also lead to distortion of their ages. The luminosities
of unresolved close binaries calculated from
their trigonometric parallaxes are higher than their
true values~[31]. Therefore, the derived ages of such
stars that have not yet reached their turn-off point
will be overestimated, while younger stars at the same
temperature will be located above this point. Our
analysis showed that this effect is manifest for stars
that have not yet passed their turn-off points, but
induces an age distortion of no more than two billion
years, which is less than the upper limit to the age
uncertainties adopted here. Nevertheless, we separated
out binary candidates in our sample using the
criterion proposed in~[32]. These comprised about
20\,\% of our stars, but the age–-metallicity diagram
constructed for the remaining single stars remained
virtually unchanged, and we therefore do not present
this diagram here.

We can also verify whether all metal-rich stars
are, in fact, young using an independent statistical
age indicator -- the stellar kinematics. We compare
the distributions of the velocities relative to the local
standard of rest for metal-rich ($[Fe/H] > 0.0$) and
metal-poor ($[Fe/H] < -0.3$) thin disk stars in our
sample from~[7], having first separated out very young
($t < 2$ billion years) and very old ($t > 5$ billion years)
stars. The Table presents the corresponding data.
Comparisons show that the parameters of the velocity
distributions for stars of the same age but different
metallicities are very similar. At the same time, old
stars display mean values and residual velocity dispersions
that are a factor of $1.5-2$ higher than those
for young stars. This suggests that the discussed
metal-rich stars do indeed have similar ages to old,
metal-poor disk stars. It is therefore important to
verify whether old, metal-rich stars have migrated
from regions closer to the Galactic center, where the
mean metallicity is higher.

\section*{INFLUENCE OF RADIAL MIGRATION OF
STARS ON THE AGE--METALLICITY RELATION}

It is shown by Roskar~et\,al. [20] (theoretically)
and Haywood~[18,\,19] (based on stars near the Sun)
that the joint action of radial migration of stars due
to relaxation effects and the negative radial metallicity
gradient in the Galactic disk lead to the large
metallicity dispersion observed for neaby stars, which,
in turn, masks any age--metallicity relation that may
intrinsically be present in these stars. According
to modern concepts, a star born from the interstellar
medium initially moves in a circular orbit. The
eccentricity of the orbit increases with time due to
interactions with the perturbations associated with
the gravitational potential of the Galaxy. However,
it is believed that the mean radii of the stellar orbits
remain virtually unchanged, reflecting the Galactocentric
distances where the stars were born [4,\,32].
Therefore, explaining the comparatively high orbital
eccentricities of supermetallic stars ($[Fe/H] > 0.2$),
Grenon~[32] proposed that they had migrated toward
the region of the Sun from distances closer to the
Galactic center. Such stars should be fairly old,
to allow time for the eccentricities and apogalactic
radii of their orbits to appreciably change. It is also
thought thatmetal-poor stars currently near the Sun,
on the contrary, were born at larger Galactocentric
distances, where the star-formation rate is lower. Let
us determine how the diagrams for stars with various
mean orbital radii will appear.

\section*{Metallicity--Age Relation for Thin-Disk Stars
Born at Different Galactocentric Distances}

To better visualize the differences between age–-
metallicity sequences for stars with different $R_{m}$ values,
Fig.~4 shows only the third-order polynomial fits
for the corresponding diagrams. (The polynomials are
prefereable to the mean points constructed in Fig.~2
in this case, since they better trace the mean dependences
when there are a small number of stars in the
sample, and are not subject to random excursions;
in the presence of sufficiently good statistics, the
polynomial fit and set of lines joining the mean points
essentially coincide.) We first separated the stars
in our sample of thin-disk stars within 70\,pc of the
Sun into two groups, with mean orbital radii larger
and smaller than the solar value. We then separated
the latter group (which contained more than twice
as many stars as the former group) into two roughly
equal groups in orbital radii divided by the value $R_{m} =
7.6$\,kpc (a maximum in the $R_{m}$ distribution for the
disk stars is observed at this value). Moreover, we
distinguished another group with approximately solar
orbital radii ($7.9 < R_{m} < 8.1$)\,kpc.

Figure~4 shows that the age–-metallicity relations
for all the groups are similar within the uncertainties.
However, with increase in the mean orbital radius,
their positions drift in a non-uniform way -- the
sequences for both groups of stars with orbital radii
smaller than the solar value become higher, while
remaining similar to each other, and even cross each
other. This behavior can be explained if stars with
increasingly higher eccentricities end up in the vicinity
of the Sun as the difference between their mean
orbital radius and the solar Galactocentric distance
increases; these stars with higher eccentricities, in
turn, include an increasingly higher fraction of stars
that are older and more metal-rich. At radii $R_{m} <
R_{\odot}$, two opposite effects act in the thin disk -- the
negative radial gradient of the metallicity and the
decrease in the metallicities of stars with increasing
age (at least over the past several billion years; see
below formore detail). At $R_{m} > R_{\odot}$, these two effects
act in the same sense. (This gives rise to a break in a
plot of the mean orbital radius versus metallicity, as is
discussed in [12,\,14].)

near the Solar Circle
\section*{Age–Metallicity Diagram for Stars Born
near the Solar Circle}

Figure~5 presents an age--metallicity diagram for
thin-disk stars located within 70\,pc of the Sun and
having mean orbital radii ($7.7 < R_{m} < 8.4$)\,kpc.
When determining the sizes of this range, we were
guided by the following considerations. First, the
variation of $[Fe/H]$ within this range due to the radial
metallicity gradient should be substantially less
than the uncertainty in this quantity. For the usually
adopted value of the gradient $grad_{R_{m}}[Fe/H] \approx-0.1$
(see, for example,~[33]), the difference in the metallicities
at the edges of this range are smaller than
the metallicity uncertainties cited in~[7]. Second,
to obtain as large a sample volume as possible, we
tried to approach as closely as possible to the maximum
in the stellar distribution for our initial sample
in mean orbital radius, observed at $R_{m} \approx 7.60$\,kpc.
We continued the interval somewhat further in the
direction of increasing $R_{m}$, to improve the balance in
the numbers of stars with larger and smaller orbital
radii, since the number of stars falls off monotonically
on either side of the maximum.
On Fig. 5, as on Fig. 2, we plot the mean metallicities
in narrow age ranges and third-order polynomial
fits to the data. A comparison of the figures indicates
that the age dependence of the metallicity remains
virtually unchanged, and the metallicity scatter for
stars of the same age has not decreased. As a result,
neither old, metal-rich ($[Fe/H] > 0.0$) nor comparatively
young, metal-poor ($[Fe/H] < -0.3$) stars from
the sample fully disappeared. The small, hollow circles
in Fig.~5 distinguish stars from the very narrow
interval ($7.9 < R_{m} < 8.1$)\,kpc, which show that,
in any section of the diagram, the number of stars
decreases roughly proportional to their initial number.
The behavior of the mean [Fe/H] values (solid curve
in Fig.~5; see also Fig.~4) is essentially independent of
the width of the interval in $R_{m}$. In particular, for stars
older than five billion years and for the total sample of
stars within 70\,kpc of the Sun, the mean values and
dispersions of the metallicities for the subsample of
stars with ($7.9 < R_{m} < 8.1$)\,kpc essentially coincide,
although the numbers of stars in these samples
differ by more than an order of magnitude. In other
words, the observed metallicity scatter is beyond the
uncertainties for any age of star, even for stars born
at the solar Galactocentric distance; however, the
mean metallicity in roughly the last five billion years
all the same changes with age. The upper and lower
envelopes in the diagram (dashed lines) also demonstrate 
an increase in the metallicity with approach
toward the current epoch.

We note one important property of the age distribution
of the disk stars that is clearly visible in the
age--metallicity diagram: the sharp increase in the
density of stars of any metallicity passing through
an age of $\approx 4.5$ billion years toward younger ages.
(Note that the distortion of the ages near three to
five billion years noted above could shift the observed
boundary for the sharp increase in the number of stars
somewhat.) The predominance of thin-disk stars with
ages $<5$ billion years was already noted earlier (see,
for example, [7,\,19,\,34]).

\section*{AGE--METALLICITY AND AGE--RELATIVE
$Mg$ ABUNDANCE RELATIONS}

Thus, we see that none of the discussed effects is
able to appreciably distort the general appearance of
the age--metallicity diagram in Fig.~2a for F--G stars
of the thin-disk subsystem located within 70\,pc of
the Sun. Nevertheless, let us trace the behavior of
the age dependences of the mean metallicity and the
metallicity dispersion constructed for the sample stars
in the range ($7.7 < R_{m} < 8.4$)\,kpc. We can see in
Fig.~6a, where these quantities were calculated in ten
narrow age bins, that the mean metallicities of thin disk
stars born at the solar Galactocentric distance
first decreases appreciably with increasing age, but
then remains constant within the uncertainties and
equal to $\langle[Fe/H]\rangle\approx -0.19$ after four to five billion
years. Figure~6b shows that the metallicity dispersion
is initially virtually independent of age and equal to $\approx 0.13$, 
but then begins to rapidly increase after four
to five billion years, reaching $\approx 0.22$ for the oldest
stars.

\section*{Use of Spectroscopic $Fe$ and $Mg$ Abundances}

Since the photometric metallicities of some stars
could be distorted by unaccounted or systematic
effects, it is necessary to also analyze the age--metallicity 
dependence based on stars with spectroscopic
metallicities, $[Fe/H]_{sp}$. The corresponding
diagram for stars within 70\,pc of the Sun is presented
in Fig.~7. Recall that the ages for these stars
were taken from~[7], where they were determined
with somewhat different spectroscopic metallicities;
however, as is shown by Fig.~1a, these differences
are within the cited uncertainties in both catalogs.
Note that unresolved binaries are clearly absent from
this sample -- otherwise, they would very likely be
known as spectroscopic binaries. since the number
of thin-disk stars in the catalog is modest, we did
not restrict the sample in the mean orbital radii of the
stars, all the moreso that, as we have demonstrated,
this leaves the character of the age dependence of the
mean metallicity essentially unchanged.

When comparing the positions of the stars in
Figs.~5 and~7a, we must allow for selection effects in
the sample with spectroscopic metallicities, including
the fact that the depth of the survey increases appreciably
with decreasing metallicity~[24]. This means
that the mean metallicity for thin-disk stars of this
last sample with ages $t > 4$ billion years is somewhat
lower than the value for the representative sample
of thin-disk stars from~[7]. Overall, the stars of the
two samples occupy the same region in the age–
metallicity diagram (Fig.~2a). (As we noted above,
this remains valid when independently determined
ages are used; Fig.~2b.) Figure~7a shows that, as
for the substantially larger sample, the mean values
of $\langle[Fe/H]_{sp}\rangle$ decrease systematically with 
increasing age, but remain unchanged within the uncertainties
at ages exceeding four to five billion years. Thus, we
conclude that features reflecting the relative metallicities
of stars with different ages in spectroscopic and
photometric age--metallicity relations coincide within
the uncertainties.

To understand the origin of the compex behavior
of the age--metallicity relation, it is also important to
trace the age dependence of the relative abundance
of $Mg$ (a representative of the $\alpha$ elements) for disk
stars. Recall that, according to modern concepts, $\alpha$
elements and a small number of Fe atoms are ejected
by type $II$ supernovae, whose evolution times before
their explosions are about 10 million years. On the
other hand, the main source of iron-peak elements
is type $Ia$ supernova explosions, which occur after
a characteristic evolution time of $\approx 1$ billion years
after the birth of the precursor star (see, for example,
[35,\,36]). Therefore, following a sudden onset of accelerated
star formation in a stellar--gaseous system,
the relative abundance of $\alpha$ elements in the newly
born stars could initially even increase somewhat;
it should then systematically decrease after about a
billion years due to the onset of an era of massive
type Ia supernovae.

The age--relative $Mg$ abundance diagram for thin disk
stars within 70\,pc of the Sun based on data from
our catalog with spectroscopic $Mg$ abundances is
presented in Fig.~7b. Indeed, an inflection is observed
in the middle of this relation. (Although the probability
of erroneously rejecting the hypothesis that
this relation can be fit by a straight line in favour of
the third-order polynomial fit is rather high due to
the small number of points, the existence of this inflection
testifies to an appreciable difference between
the mean ratios $\langle[Mg/Fe]\rangle$ for stars younger than
two billion years ($\langle[Mg/Fe]\rangle = 0.04 \pm 0.01$) and older
than four billion years ($\langle[Mg/Fe]\rangle = 0.11 \pm 0.01$).)
As a result, the mean relative abundance of $Mg$ in
stars of the thin disk, which is fairly high in the
initial stages of its formation ($\langle[Mg/Fe]\rangle \approx 0.10$) 
and is independent of the age within the uncertainties,
began to appreciably more rapidly decrease with approach
toward the current epoch starting several billion
years ago. Unfortunately, the insufficient number
of stars with spectroscopic relative $Mg$ abundances
and large age uncertainties hinders our ability to trace
the behavior of the age dependence of this abundance
in detail, and to statistically confidently identify the
epoch of the onset of the more rapid decrease in the
ratio $\langle[Mg/Fe]\rangle$. However, comparison of the 
age--metallicity and age--relative $Mg$ abundance diagrams
shows that the decrease in the mean relative $Mg$
abundance in thin-disk stars appears to have begun
approximately three to four billion years ago, somewhat
later than the increase in the mean metallicity.
The observed decrease in $[Mg/Fe]$ right to the current
epoch testifies that, following the sudden burst of star
formation, the rate of this decrease slowed somewhat,
but always remained higher than before this burst.

THE CHEMICAL COMPOSITION IN THE
THIN DISK%
\section*{SCENARIOS FOR THE EVOLUTION OF
THE CHEMICAL COMPOSITION IN THE
THIN DISK}

Thus, if we suppose that the mean orbital radii
of stars indeed reflects the Galactocentric distances
of their birth places, both the age [Fe/H] and age--$[Mg/Fe]$ 
diagrams testify that, during the first several
billion years of formation of the thin-disk subsystem,
the interstellar material was, on average, fairly rich
in heavy elements ($\langle[Fe/H]\rangle \approx -0.2$), while the mean
relative $Mg$ abundance was only slightly higher than
the solar value ($\langle[Mg/Fe]\rangle \approx 0.1$). The dispersion of
the heavy-element abundances were fairly large. The
reason for this constancy of the mean heavy-element
abundance and its large dispersion in the newly born
stars is likely a low star-formation rate, together with
a continuous raining of metal-poor interstellar matter
from outer parts of the Galaxy onto the disk. However,
approximately four to five billion years ago, the
mean metallicity began to systematically increase,
while the mean relative $Mg$ abundance began to decrease
slightly later. This probably occurred due to a
sudden increase in the star-formation rate in the thin
disk, which is confirmed by the substantial increase
in the number of stars beginning approximately from
this age. As a result of continuous mixing of the
interstellar medium, the high dispersion of the metallicity
at the epoch of formation of the first thin-disk
stars ($\sigma_{[Fe/H]}\approx 0.22$) systematically decreases during
the first several billion years of the formation of this
subsystem. However, four to five billion years ago, the
onset of a burst of star formation stopped this process,
and the metallicity dispersion subsequently remained
constant right to the current epoch ($\sigma_{[Fe/H]}\approx 0.13$).

In this picture, like the high mean metallicity of
stars in early stages of the formation of the thin disk
subsystem, the existence of metal-rich old stars
can be explained by the fact that, in a strongly inhomogeneous
interstellar medium in terms of its metal
content, star formation proceeds more intensively in
clouds with enhanced heavy-element contents. This
simultaneously testifies to the fact that the main enrichment
in metals in the Galaxy occurred in previous
stages in its evolution, whereas this enrichment
was modest in the thin-disk stage and has
been appreciable only over the last four to five billion
years. This provides further support for the existence
of a prolonged interval in which star formation was
suppressed (an active phase in the evolution of the
Galaxy), between the formation of the main mass of
stars in the thick and thin Galactic disks [29,\,30].

It is interesting that the epoch $\approx 4$ billion years ago
(redshifts of $z \approx 0.4$) appears to have been special in
a more general sense in the Universe. A massive
''transformation'' of spiral into lenticular galaxies is
observed in galaxy clusters roughly at this epoch:
while spiral galaxies dominate ($70\,\%$) in clusters at
$z > 0.5$, they almost disappear after $z < 0.4$, when
the dominant population becomes lenticular galaxies
[37--39]. At approximately the same time, nearly all
of the most massive spheroidal and elliptical dwarf
galaxies in the Local Group displayed bursts of star
formation (see, for example, [40,\,41]). These facts
suggest that our Galaxy, which had previously been
isolated, became one of the central galaxies in the
Local Group about four billion years ago. The outer
gaseous halo of the Galaxy probably fell into its center
and disk due to tidal processes. (No fairly massive
gaseous halo is currently observed in our Galaxy,
although it is possible that such a halo existed before
the Galaxy became a member of the Local Group.) It
may be that precisely this event provoked an amplification
of star formation in the thin disk of the galaxy,
and also led to a cessation of the infall of metal-poor
gas. This scenario is also in good agreement with
theoretical modeling of the ''clustering'' of galaxies
in groups of various mass, according to which the
number of galaxy clusters with masses comparable
to the Local Group reaches saturation at $z\approx 0.4$ 
[42,\,Fig.~4].

This picture overall differs from those considered
previously (see the Introduction), although all its features
have been noted separated in earlier studies.
Our preference for this model is due to the fact that,
first, we have constructed the age--metallicity diagram
here for the first time based on an essentially
general collection of single F--G dwarfs located within
70\,pc of the Sun; second, the number of stars
belonging to the thick disk has been minimized in our
sample; and, third, the results have been refined based
on stars born at the solar Galactocentric distance.
Unfortunately, the insufficient depth of the Geneva.
Copenhagen survey and limited trustworthiness of
the stellar ages in the range 3.5 billion years hinder
us from composing a sample large enough to trace the
epochs of the onset of the sharp increase in the star 
formation rate at various distances from the Galactic
center.

\section*{ACKNOWLEDGMENTS}

The authors thank O.K. Sil'chenko for suggesting
the possible relation between internal processes in our
Galaxy and the global clustering of galaxies in the
Universe.
This work was partially supported by the Russian Federation for Basic 
Research (project 11-02-00621-a), the Ministry of Education 
(project P-685), and the Federal Agency for Science and Innovation 
(project 02.740.11.0247).


\newpage

\begin{figure*}
\centering
\includegraphics[angle=0,width=0.70\textwidth,clip]{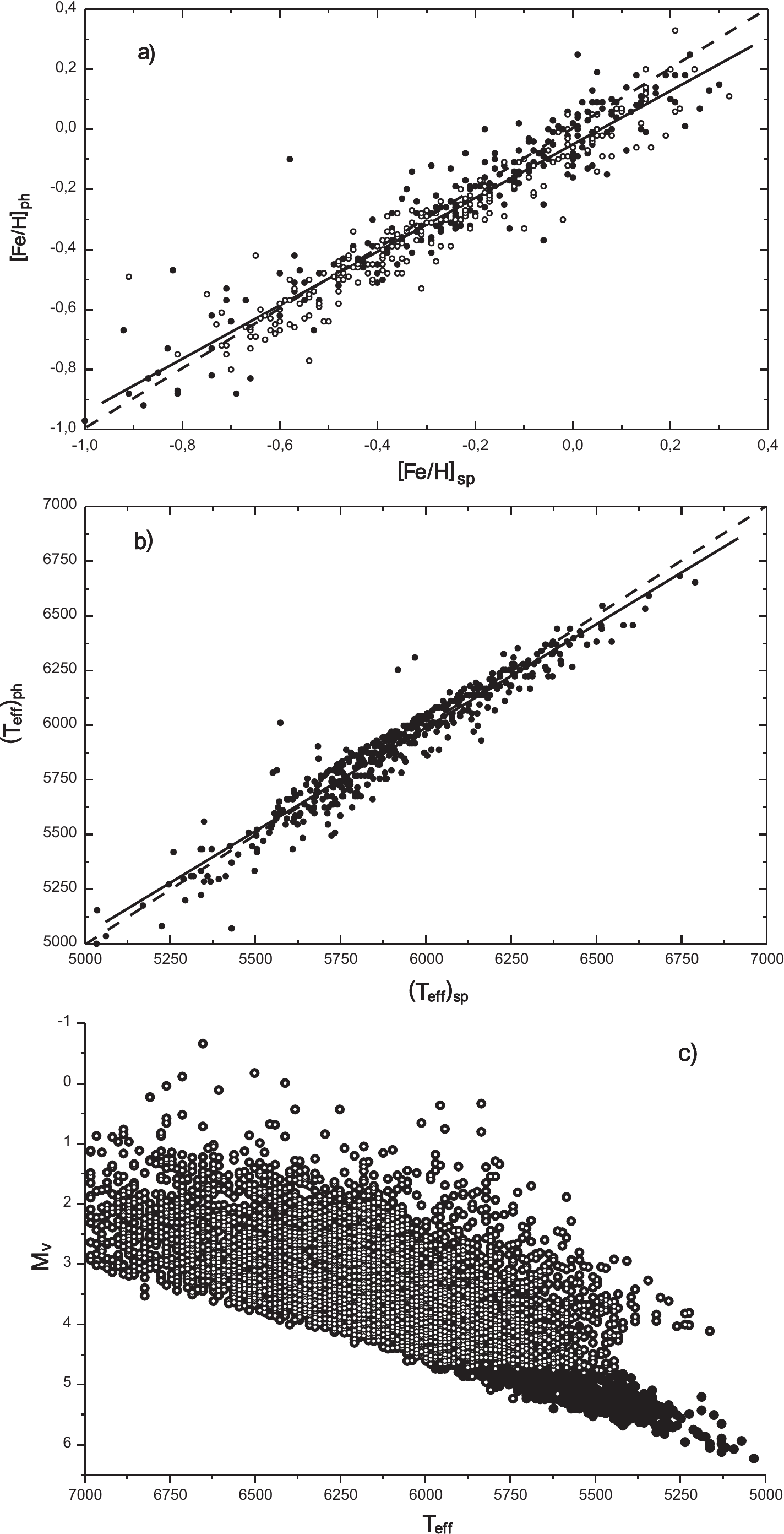}
\caption{(a) Comparison of photometric and spectroscopic 
metallicities. The hollow circles denote stars with ages 
exceeding five billion years. (b) Comparison of photometric 
and spectroscopic effective temperatures. The bold lines in (a) 
and (b) correspond to a perfect agreement between the compared 
quantities, while the thin lines show the results of linear fits. 
(c) Plot of effective temperature versus absolute magnitude. 
Small symbols show all stars from the input catalog, while 
hollow circles show stars with age uncertainties $\epsilon(t) < 3$ 
billion years.}
\label{fig1}
\end{figure*}

\newpage

\begin{figure*}
\centering
\includegraphics[angle=0,width=0.90\textwidth,clip]{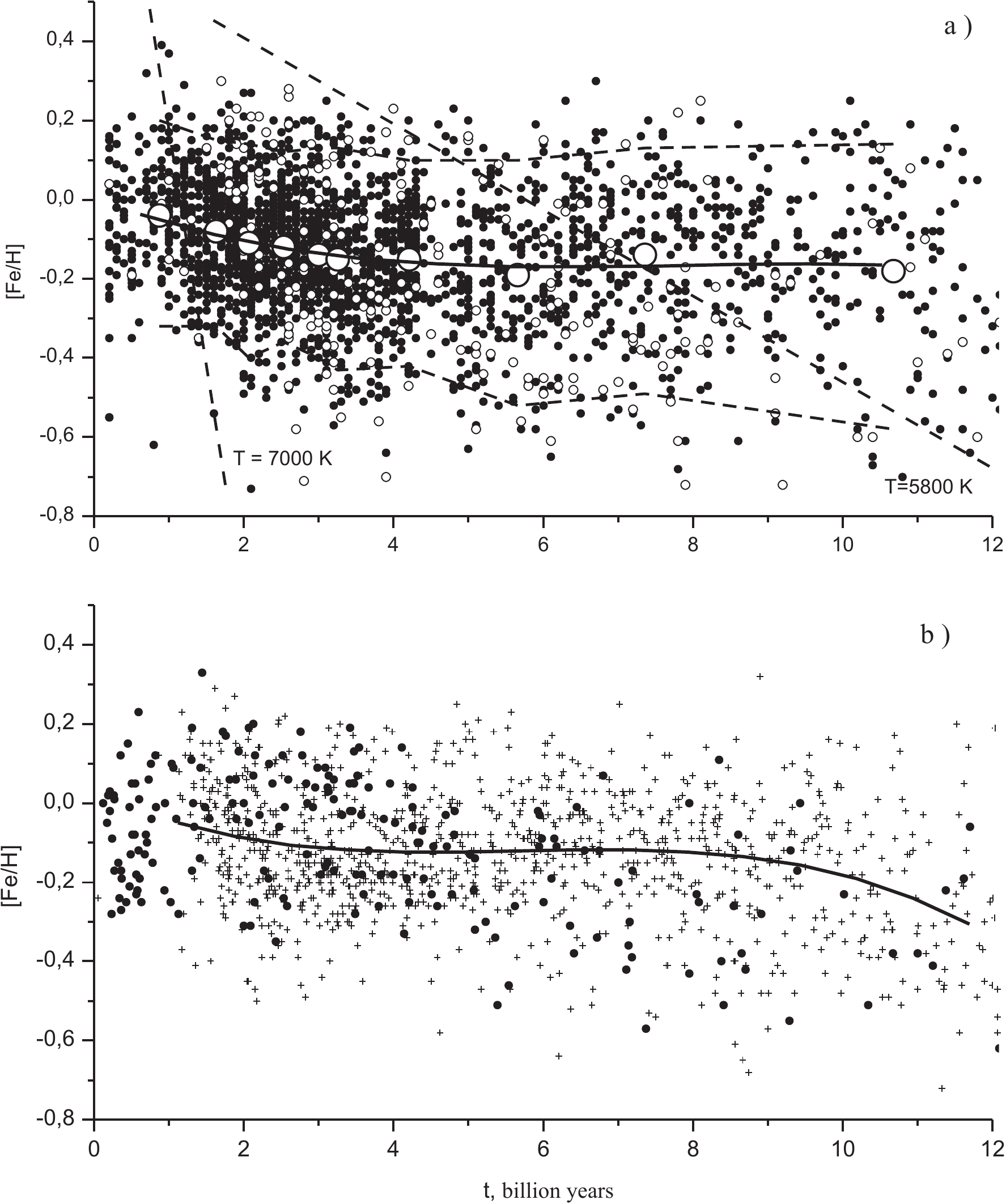}
\caption{Age--metallicity diagram for thin-disk stars with 
$\epsilon(t) < \pm 3$ billion years located within 70\,pc 
of the Sun, from the catalogs (a)~[7] and (b)~[5] 
(crosses) and~[11] (circles). The smooth curves are third-order 
polynomial fits of the $[Fe/H]$--age dependences for all 
stars in the samples of~[7] (a) and~[5] (b). The large hollow circles 
in (a) show the mean metallicities of stars within narrow age 
ranges; the broken dashed lines show the upper and lower 5\,\% 
envelopes; the sloped dashed lines show theoretical isotherms for 
$T_{eff} = 7000$ and 5800\,K; the small hollow circles show stars with 
spectroscopic $[Fe/H]$ values from~[24] and ages from~[7]. The small 
number of stars with ages greater than 12 billion years are not shown 
on the diagram.}
\label{fig2}
\end{figure*}

\newpage

\begin{figure*}
\centering
\includegraphics[angle=0,width=0.96\textwidth,clip]{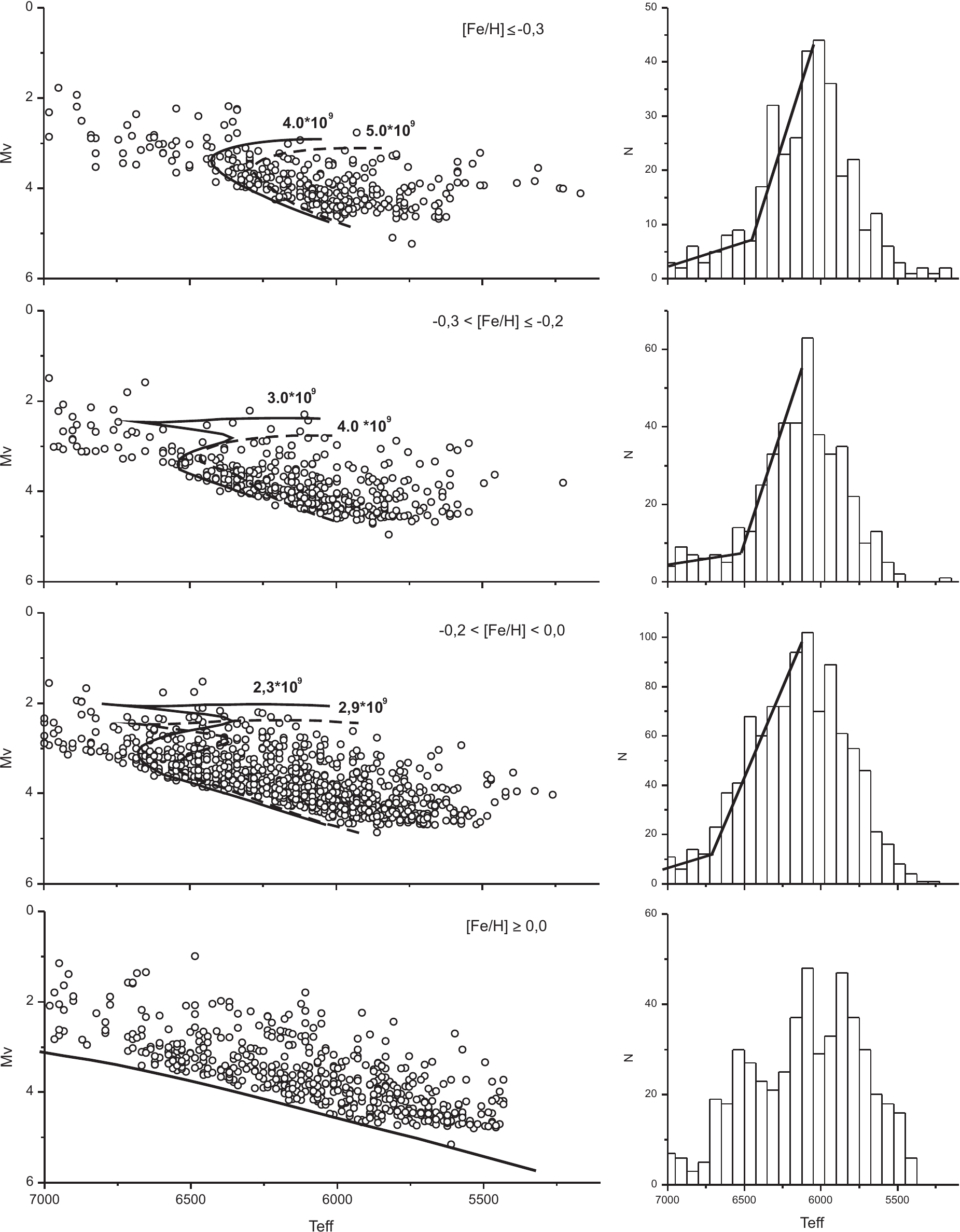}
\caption{Plot of $T_{eff}$ versus $M_{V}$ for stars within 70\,pc of 
the Sun in four narrow metallicity ranges (left) and temperature
distributions for the same stars (right). The line segments on the 
historgrams underscore the positions of sharp breaks in
their high-temperature intervals. Theoretical isochrones from~[17] 
whose turn-off points correspond to the positions of the
breaks in the histograms are shown in the left-hand plots 
(solid curves); isochrones for a higher age are shown to the right of
these for comparison (dashed curves). The ages of the isochrones are 
indicated (in years).}
\label{fig3}
\end{figure*}

\newpage

\begin{figure*}
\centering
\includegraphics[angle=0,width=0.96\textwidth,clip]{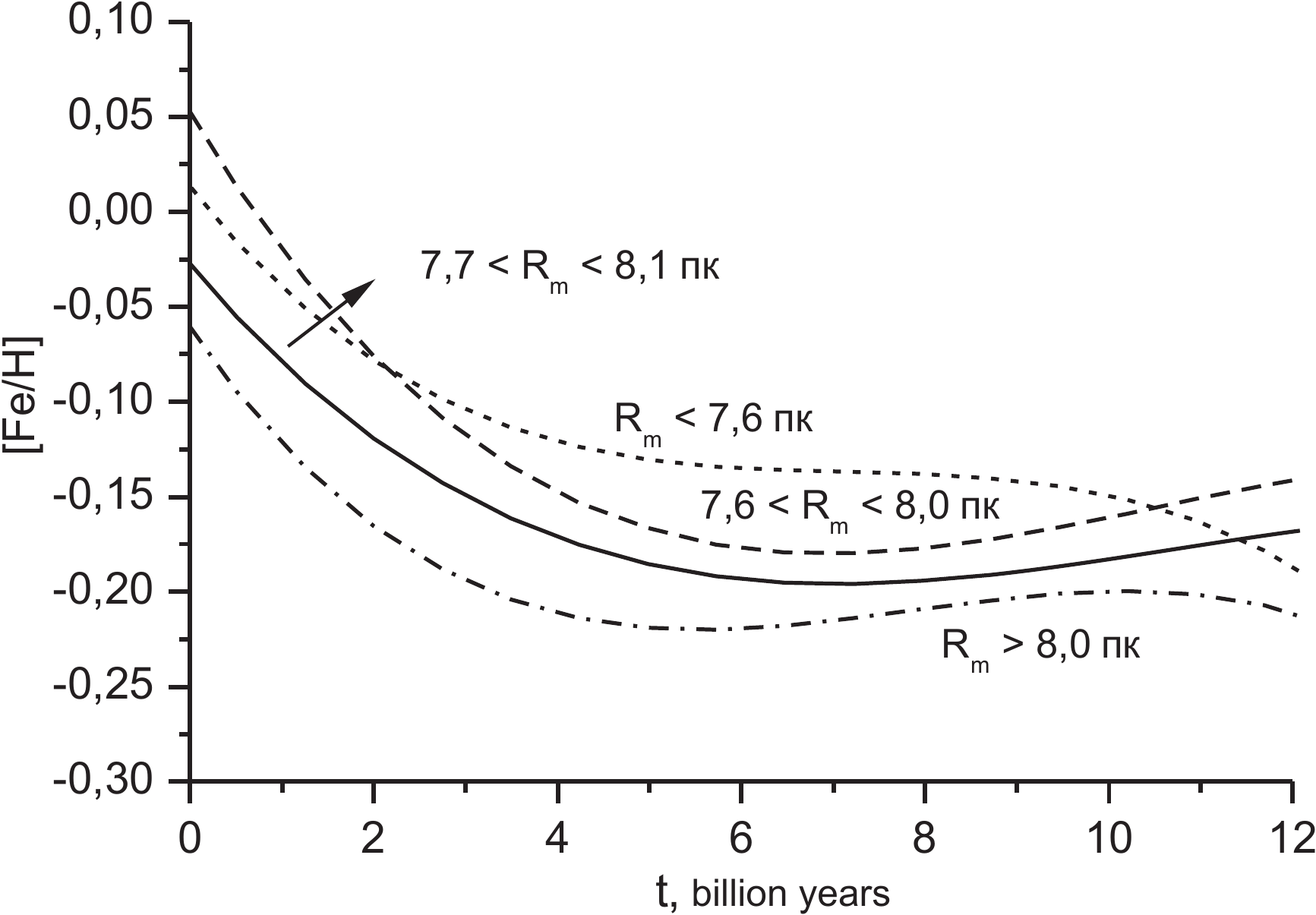}
\caption{Dependences fit to the metallicity--age relation for thin-disk 
stars located within 70\,pc of the Sun and having mean
orbital radii $R_{m} < 7.6$\,kpc, ($7.6  < R_{m} < 8.0$)\,kpc, 
($7.9 < R_{m} < 8.1$)\,kpc, and $R_{m} > 8.0$\, kpc. 
The drift in the position of the fit dependence with increase 
in the mean orbital radius, while preserving the general form of 
the trends, is clearly visible.}
\label{fig4}
\end{figure*}

\newpage

\begin{figure*}
\centering
\includegraphics[angle=0,width=0.96\textwidth,clip]{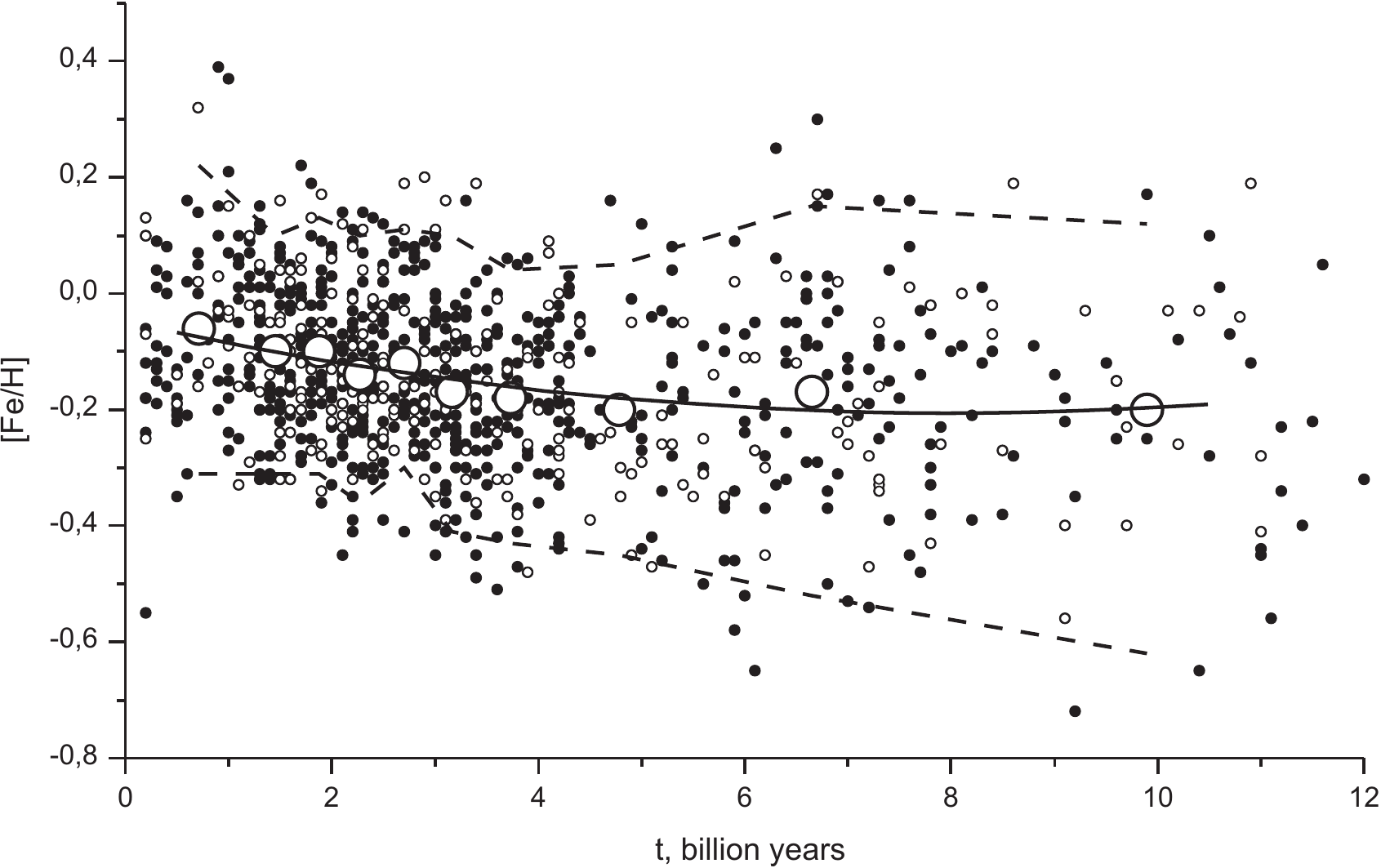}
\caption{Age--metallicity diagram for thin-disk stars located within 
70\,pc of the Sun and having mean orbital radii of
($7.7 < R_{m} < 8.4$)\,kpc. The notation is the same as in Fig. 2a. 
The small, hollow circles show stars with mean orbital
radii ($7.9 < R_{m} < 8.1$)\,kpc.}
\label{fig5}
\end{figure*}

\newpage

\begin{figure*}
\centering
\includegraphics[angle=0,width=0.96\textwidth,clip]{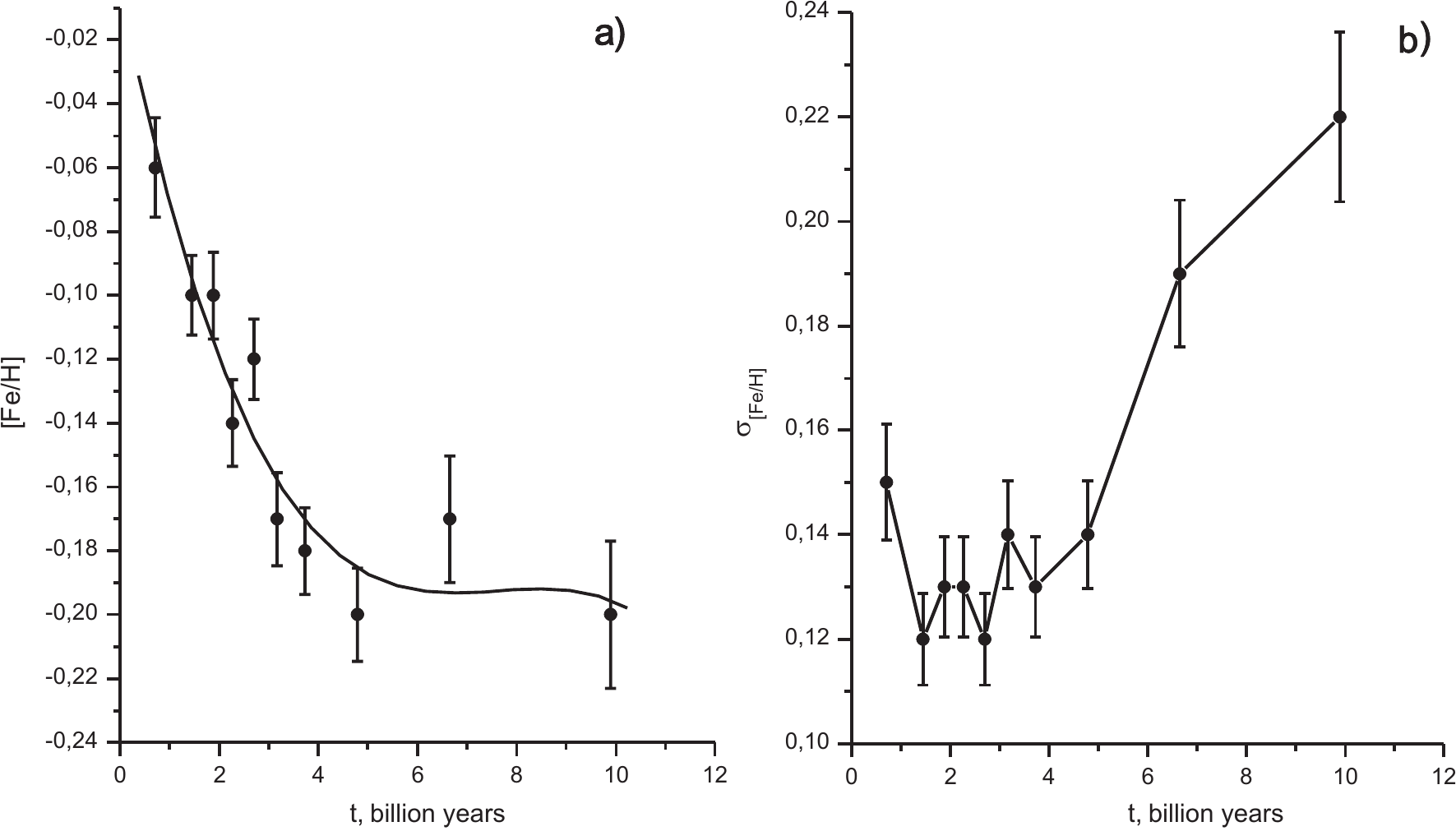}
\caption{Age dependences of the (a) mean value and (b) dispersion 
of the metallicity for thin-disk stars within 70\,pc of the
Sun with mean orbital radii ($7.7 < R_{m} < 8.4$)\,kpc. The bars 
show the uncertainties. The curve in the left panel shows a
third-order polynomial fit.}
\label{fig6}
\end{figure*}

\newpage

\begin{figure*}
\centering
\includegraphics[angle=0,width=0.96\textwidth,clip]{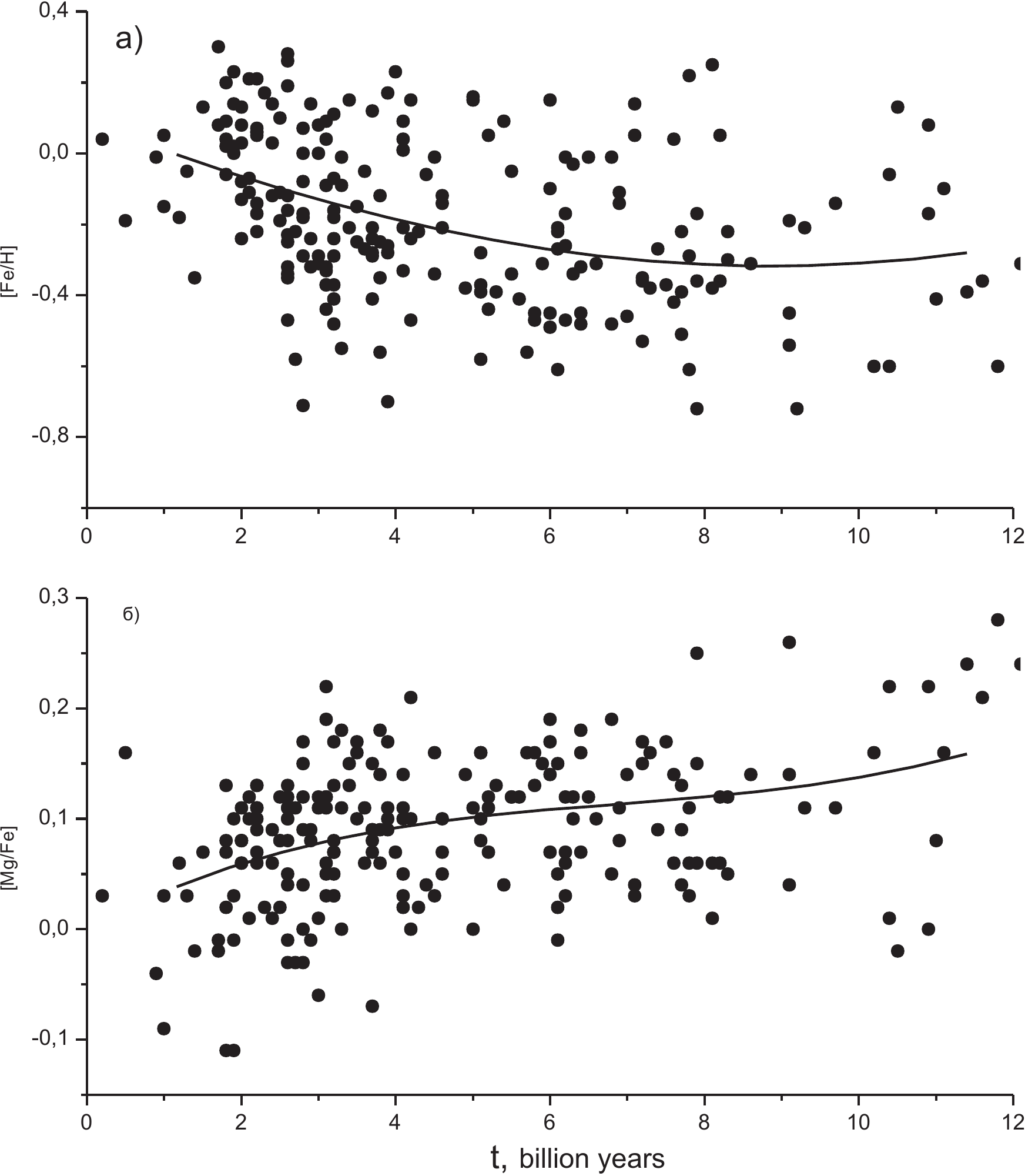}
\caption{(a) Age--$[Fe/H]$ and (b) age--$[Mg/Fe]$ diagrams for 
thin-disk stars located within 70 pc of the Sun with spectroscopic
$Fe$ and $Mg$ abundances from~[24] and age uncertainties 
$\epsilon(t) < \pm 3$ billion years. The curves show 
third-order polynomial fits.}
\label{fig7}
\end{figure*}

\end{document}